\documentclass[preprint,showpacs,preprintnumbers,amsmath,amssymb]{revtex4}
\usepackage{mathrsfs}
\usepackage{graphics}
\usepackage{dcolumn}
\usepackage{bm}

\begin{document}

\title{Representation theory for vector electromagnetic beams}

\author{Chun-Fang Li\footnote{On leave of absence from the Dipartimento di Fisica, Universit\`{a} Roma Tre, Via della Vasca Navale 84, I-00146
Rome, Italy. Email address: cfli@shu.edu.cn}}

\affiliation{Department of Physics, Shanghai University, 99 Shangda Road, 200444
Shanghai, China}

\affiliation{State Key Laboratory of Transient Optics and Photonics, Xi'an Institute of Optics and
Precision Mechanics of CAS, 710119 Xi'an, China}


\begin{abstract}

A representation theory of finite electromagnetic beams in free space is formulated by factorizing
the field vector of the plane-wave component into a $3 \times 2$ mapping matrix and a 2-component
Jones-like vector. The mapping matrix has one degree of freedom that can be described by the
azimuthal angle of a fixed unit vector with respect to the wave vector. This degree of freedom
allows us to find out such a beam solution in which every plane-wave component is specified by the
same fixed unit vector $\mathbf{I}$ and has the same normalized Jones-like vector. The angle
$\theta_I$ between the fixed unit vector and the propagation axis acts as a parameter that
describes the vectorial property of the beam. The impact of $\theta_I$ is investigated on a beam
of angular-spectrum field scalar that is independent of the azimuthal angle. The field vector in
position space is calculated in the first-order approximation under the paraxial condition. A
transverse effect is found that a beam of elliptically-polarized angular spectrum is displaced
from the center in the direction that is perpendicular to the plane formed by the fixed unit
vector and the propagation axis. The expression of the transverse displacement is obtained. Its
paraxial approximation is also given.

\end{abstract}

\pacs{41.20.Jb, 02.10.Yn, 42.25.Ja}          
\maketitle


\section{Introduction}

The representation formalism and propagation characteristics of an electromagnetic beam in free
space has drawn much attention \cite{Lax-LM, Pattanayak-A, Davis, Davis-P1, Davis-P2, Gori-GP,
Durnin-ME, Jordan-H, Tovar-C, Enderlein-P, Seshadri, Li1} after the advent of masers and lasers
\cite{Green-W, Kogelnik, Kogelnik-L}. It was shown \cite{Lax-LM} that a linearly polarized beam
solution is not compatible with the Maxwell equations, because it does not satisfy the
transversality condition. It was also shown \cite{Pattanayak-A} that the state of polarization is
not a global property of a finite beam. Rather it is local and changes on propagation. In an
attempt to describe the vectorial property of a beam, a unit vector was once introduced and was
first supposed \cite{Pattanayak-A, Davis} to be perpendicular to the propagation axis. Later on,
it was further pointed out \cite{Davis-P1, Davis-P2} that the unit vector can also be parallel to
the propagation axis. Under the paraxial condition, the beam in the former case is uniformly
polarized, and the beam in the latter case is now known as the cylindrical vector beam
\cite{Youngworth-B1, Li1}. The conversion from uniformly polarized beams to cylindrical vector
beams has been experimentally realized \cite{Tidwell-FK, Stalder-S, Youngworth-B1, Bomzon-BKH,
Ren-LW}.

Recently, there appeared a controversy \cite{Onoda-MN1, Bliokh-B1, Bliokh-B2} over the physical
properties of the light beams that were proposed to investigate the Imbert-Fedorov effect, a
transverse displacement of a reflected \cite{Fedorov, Imbert, Pillon-GG, Fedoseev1, Fedoseev2} or
a transmitted \cite{Schilling, Fedoseev1, Fedoseev2, Onoda-MN2, Hosten-K} beam taking place at a
dielectric interface. On the one hand, Onoda et. al. \cite{Onoda-MN1} disagreed with Bliokh and
Bliokh \cite {Bliokh-B1} on their incident beam. On the other hand, Bliokh and Bliokh
\cite{Bliokh-B2} found that the physical properties of Onoda's incident beam \cite{Onoda-MN1}
depend on the ``incidence angle''. Such a controversy concerns in fact the description of the
vectorial property of a finite beam.

In this paper, I will show that the vectorial property of a finite beam can be described by a
parameter, the angle between a unit vector and the propagation axis. This is achieved by
factorizing the field vector of a plane wave into a $3 \times 2$ mapping matrix (MM) and a
2-component Jones-like vector \cite{Li2, Li1, Jones} and investigating the degree of freedom of
the MM. It is shown in Section \ref{degree of freedom of MM} that the MM can not be determined
uniquely by the transversality condition. It has one degree of freedom. The degree of freedom can
be represented by the azimuthal angle of a fixed unit vector with respect to the wave vector. The
idea of MM is generalized in Section \ref{representation formalism} to a finite beam. For an
arbitrary electromagnetic beam, each plane wave component may have its own MM and Jones-like
vector. But the degree of freedom of the MM allows us to find out such a beam solution in which
every plane wave component is specified by the same fixed unit vector and has the same normalized
Jones-like vector. In this case, the beam as a whole has its own MM, which maps the normalized
Jones-like vector to the field vector. The case of a transverse unit vector corresponds to the
representation formalism discussed in Refs. \cite{Pattanayak-A, Davis}, and the case of a
longitudinal unit vector corresponds to the representation formalism discussed in Refs.
\cite{Davis-P1, Davis-P2}. The impact of the angle between the unit vector and the propagation
axis is investigated in Section \ref{impact}. The field vector in the first-order approximation
under the paraxial condition is calculated for an angular-spectrum field scalar that is
independent of the azimuthal angle. The controversy over the incident beams of Refs.
\cite{Onoda-MN1} and \cite{Bliokh-B1} is resolved. A transverse effect is also found that a beam
of elliptically polarized angular spectrum is displaced from the center in the direction that is
perpendicular to the plane formed by the unit vector and the propagation axis. The origin of this
effect is discussed. Conclusions and remarks are given in Section \ref{conclusions}.

\section{\label{degree of freedom of MM} Degree of freedom of the mapping matrix}

\subsection{Mapping matrix and the existence of one degree of freedom}

In a source-free position space, the electric-field vector $\mathbf{F}(\mathbf{x})$ of a
monochromatic finite electromagnetic beam satisfies the transversality condition,
\begin{equation} \label{transversality}
\nabla \cdot \mathbf{F}(\mathbf{x})=0.
\end{equation}
The plane-wave angular-spectrum expression of the field vector, varying according to $\exp(-i
\omega t)$ with the time, can be written as
\begin{equation}\label{electric field of a beam}
\mathbf{F}(\mathbf{x})= \frac{1}{2\pi} \int \int_{k_x^2+k_y^2 \leq k^2} \mathbf{f} (k_x,k_y)
\exp(i \mathbf{k} \cdot \mathbf{x}) d k_x d k_y,
\end{equation}
where $\mathbf{k}= k_x \mathbf{e}_x+ k_y \mathbf{e}_y+ k_z \mathbf{e}_z \equiv \left(
\begin{array}{c} k_x \\ k_y \\ k_z \end{array} \right)$ is the wave vector satisfying $k_x^2+ k_y^2+
k_z^2= k^2$, $\mathbf{f}= \left( \begin{array}{c} f_x \\ f_y \\ f_z
\end{array} \right)$ is the electric-field vector of the angular spectrum, and $\mathbf{e}_j$ ($j=x,y,z$) is the unit vector of the
$j$-axis. According to the transversality condition (\ref{transversality}), $\mathbf{f}$ has only
two mutually orthogonal polarization states, each of them being orthogonal to $\mathbf{k}$.
Denoting respectively by $p$ and $s$ the two orthogonal linearly-polarized states, we can
decompose $\mathbf{f}$ as
\begin{equation} \label{decomposition}
\mathbf{f}= \mathbf{f}_p+ \mathbf{f}_s= f_p \mathbf{p}+ f_s \mathbf{s},
\end{equation}
where $f_p$ and $f_s$ are respectively the $p$- and $s$-polarized complex amplitudes constituting
a 2-component Jones-like vector \cite{Li2, Li1, Jones}
\begin{equation} \label{Jones vector}
\tilde{f} \equiv \left(
                       \begin{array}{c} f_p \\ f_s \end{array}
                 \right),
\end{equation}
$\mathbf{p}$ and $\mathbf{s}$ are respectively the $p$- and $s$-polarized unit vectors. To be
clear, we assume in this paper that both $\mathbf{p}$ and $\mathbf{s}$ are real unit vectors. We
can always do this as one may see from Eq. (\ref{decomposition}). They satisfy
\begin{equation} \label{perpendicular to k}
\left\{
\begin{array}{rcl}
\mathbf{p} \cdot \mathbf{k} & = & 0, \\
\mathbf{s} \cdot \mathbf{k} & = & 0,
\end{array}
\right.
\end{equation}
as well as
\begin{equation} \label{perpendicular to each other}
\left\{
\begin{array}{rcl}
\mathbf{p} \cdot \mathbf{p} & = & 1, \\
\mathbf{s} \cdot \mathbf{s} & = & 1, \\
\mathbf{p} \cdot \mathbf{s} & = & 0.
\end{array}
\right.
\end{equation}
Eq. (\ref{decomposition}) means that the 3-component field vector is an element of a 2D space,
rather than an element of a 3D space. Since the space of 2-component Jones-like vectors is a 2D
space, Eq. (\ref{decomposition}) defines in fact a mapping from the Jones-like-vector space to the
3-component 2D space. In order to describe this mapping, we change the form of Eq.
(\ref{decomposition}) into
\begin{equation} \label{factorization}
\mathbf{f}= m \tilde{f},
\end{equation}
where
\begin{equation} \label{MM}
m= \left(
         \begin{array}{cc}
                p_x & s_x \\ p_y & s_y \\ p_z & s_z
         \end{array}
   \right)
\end{equation}
is the $3 \times 2$ MM.

Any MM, the column vectors of which satisfy Eqs. (\ref{perpendicular to k}) and
(\ref{perpendicular to each other}), guarantees that the field vector $\mathbf{f}$ given by Eq.
(\ref{factorization}) satisfies the transversality condition whatever the Jones-like vector
$\tilde{f}$ may be. But there are only five equations to determine the six unknown elements of the
MM. This shows that after the transversality condition is taken into account, the MM still has one
degree of freedom. That is to say, the transversality condition itself is not sufficient to
determine an electromagnetic wave from a given Jones-like vector.

\subsection{Description of the degree of freedom and its unique role}

\begin{figure}[ht]
\includegraphics{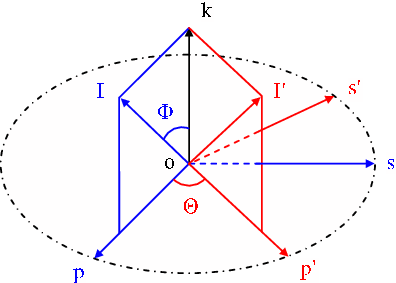}
\caption{(Color online) The degree of freedom of the MM is denoted by the azimuthal angle $\Theta$
of $\mathbf{I}$ with respect to $\mathbf{k}$.} \label{orientation}
\end{figure}

It is well known \cite{Green-W, Pattanayak-A, Davis-P2, Onoda-MN1} that if the unit vectors
$\mathbf{p}$ and $\mathbf{s}$ are defined from the wave vector $\mathbf{k}$ in terms of an
arbitrary fixed real unit vector $\mathbf{I}$ as
\begin{equation} \label{triad}
\left\{
\begin{array}{rcl}
\mathbf{p} & = & \mathbf{s} \times \frac{\mathbf{k}}{k}, \\
\mathbf{s} & = & \frac{\mathbf{k} \times \mathbf{I}}{|\mathbf{k} \times \mathbf{I}|},
\end{array}
\right.
\end{equation}
then they satisfy Eqs. (\ref{perpendicular to k}) and (\ref{perpendicular to each other}). One
might conclude that the degree of freedom is the unit vector $\mathbf{I}$. This is obviously not
true, because we need two independent variables to determine the orientation of a real unit vector
in a 3D space. But we can indeed use the real unit vector $\mathbf{I}$ to denote the degree of
freedom somehow. To show this, let us look at a particular wave vector $\mathbf{k}$ and the unit
vectors $\mathbf{p}$ and $\mathbf{s}$ that are defined by Eqs. (\ref{triad}) in terms of a fixed
unit vector $\mathbf{I}$ as is schematically depicted in Fig. \ref{orientation}. It can be seen
from this figure that the rotation of $\mathbf{I}$ around $\mathbf{k}$ by changing the azimuthal
angle $\Theta$ of $\mathbf{I}$ with respect to $\mathbf{k}$ alters the orientation of $\mathbf{p}$
and $\mathbf{s}$. On the other hand, the rotation of $\mathbf{I}$ around $\mathbf{s}$ by changing
the polar angle $\Phi$ of $\mathbf{I}$ with respect to $\mathbf{k}$ does not alter the orientation
of $\mathbf{p}$ and $\mathbf{s}$. So it is the azimuthal angle $\Theta$ that plays the role of the
degree of freedom and uniquely determines the MM. For a particular wave vector, different values
of $\Theta$ represent different mapping matrices, and vice versa. The rotation of $\mathbf{I}$
around $\mathbf{s}$ forms a group $G(\Theta)$ that corresponds to one single value of the degree
of freedom.

Based on the above description of the MM's degree of freedom, it is concluded that different
mapping matrices map a given Jones-like vector to different field vectors.

\section{\label{representation formalism} Representation formalism for a finite beam}

On factorizing the field vector of a plane wave into the MM and the Jones-like vector, we have
identified one degree of freedom of the MM and shown that it can be described by the azimuthal
angle of a fixed unit vector with respect to the wave vector. But we have not made any
requirements on the Jones-like vector. A finite beam consists of an infinite number of plane
waves. For an arbitrary beam, each plane-wave component may have its own MM and Jones-like vector.
But the degree of freedom of the MM allows us to find out such a kind of beams in which every
plane-wave component is specified by the same fixed unit vector and have the same normalized
Jones-like vector. In the following I will put forward the representation formalism of those
beams.

Assume that the fixed unit vector $\mathbf{I}$ that is common to all the plane-wave components
lies in the plane $zox$ and makes an angle $\theta_I$ with the $z$-axis, the propagation axis,
\begin{equation} \label{vector u}
\mathbf{I} (\theta_I)= \mathbf{e}_z \cos \theta_I+ \mathbf{e}_x \sin \theta_I,
\end{equation}
so that the MM of the plane-wave component takes the form of
\begin{equation} \label{form of MM}
m= \frac{1}{k |\mathbf{k} \times \mathbf{I}|}
   \left(
         \begin{array}{cc}
                 (k_y^2+k_z^2) \sin \theta_I- k_z k_x \cos \theta_I &  k k_y \cos \theta_I \\
                -k_y (k_z \cos\theta_I+ k_x \sin\theta_I)           &  k (k_z \sin\theta_I- k_x \cos\theta_I) \\
                 (k_x^2+k_y^2) \cos \theta_I- k_z k_x \sin \theta_I & -k k_y \sin \theta_I
         \end{array}
   \right),
\end{equation}
where
\begin{equation} \label{normalization factor}
|\mathbf{k} \times \mathbf{I}|= [k^2- (k_z \cos \theta_I+ k_x \sin \theta_I)^2]^{1/2}.
\end{equation}
It should be noted that a given $\theta_I$ corresponds to the same value of the MM's degree of
freedom for each of those wave vectors that also lie in the $zox$ plane as was shown before in
Fig. \ref{orientation}. Such a feature of $\theta_I$ will produce a very interesting transverse
effect as will be shown in Section \ref{impact}.

As discussed before, the field vector $\mathbf{f}$ of the angular spectrum is factorized by Eq.
(\ref{factorization}) into the MM (\ref{form of MM}) and the Jones-like vector,
\begin{equation} \label{separation}
\tilde{f}= \tilde{\alpha} f,
\end{equation}
where $\tilde{\alpha}= \left( \begin{array}{c} \alpha_p\\\alpha_s \end{array} \right)$ is the
normalized Jones-like vector describing the polarization state of the angular spectrum, $\alpha_p$
and $\alpha_s$ are complex numbers satisfying the normalization condition,
\begin{equation} \label{normalization condition}
|\alpha_p|^2+|\alpha_s|^2= 1,
\end{equation}
and $f$ is referred to as the field scalar of the angular spectrum. If $\tilde{\alpha}$ is common
to all the plane-wave components, the electric-field vector (\ref{electric field of a beam}) takes
the following factorized form,
\begin{equation} \label{field vector of beam in MM}
\mathbf{F}(\mathbf{x})= M(\mathbf{x}) \tilde{\alpha},
\end{equation}
where
\begin{equation} \label{MM of beam}
M(\mathbf{x})= \frac{1}{2\pi} \int \int m f \exp(i \mathbf{k} \cdot \mathbf{x}) d k_x d k_y
\end{equation}
is the MM for the beam, and the integration limit ${k_x^2+k_y^2 \leq k^2}$ is omitted for brevity.
Eq. (\ref{field vector of beam in MM}) states that the field vector in position space is the
result of the mapping of the normalized Jones-like vector by the MM (\ref{MM of beam}). Given a
normalized Jones-like vector, altering only $\theta_I$ that is involved in the MM (\ref{MM of
beam}) will change the vectorial property of the beam. That is to say, $\theta_I$ behaves as a
parameter that describes the vectorial property of the beam. It is not necessarily equal to
$\frac{\pi}{2}$ \cite{Pattanayak-A, Davis} nor equal to $0$ \cite{Davis-P1, Davis-P2}.

In a circular cylindrical system with the $z$-axis being the symmetry axis, the integral (\ref{MM
of beam}) is changed into
\begin{equation} \label{MM in cylindrical system}
M(\mathbf{x})= \frac{1}{2 \pi} \int_0^k \int_0^{2 \pi}  m f  \exp(i \mathbf{k} \cdot \mathbf{x})
k_{\rho} dk_{\rho} d\varphi,
\end{equation}
where $\mathbf{x}=\mathbf{r}+ z \mathbf{e}_z$, $\mathbf{r}= r \mathbf{e}_r= \mathbf{e}_x r \cos
\phi+ \mathbf{e}_y r \sin \phi$, $\mathbf{e}_r$ and $\mathbf{e}_{\phi}$ are respectively the unit
vectors in the radial and azimuthal directions in position space; correspondingly, $\mathbf{k}=
\mathbf{k}_{\rho}+ k_z \mathbf{e}_z$, $\mathbf{k}_{\rho}= k_{\rho} \mathbf{e}_{\rho}= k_x
\mathbf{e}_x+ k_y \mathbf{e}_y$, $k_x= k_{\rho} \cos \varphi$, $k_y= k_{\rho} \sin \varphi$, $k_z=
(k^2- k_{\rho}^2)^{1/2}$, $\mathbf{e}_{\rho}$ and $\mathbf{e}_{\varphi}$ are respectively the unit
vectors in the radial and azimuthal directions in wave-vector space.

\section{\label{impact} Impact of angle $\theta_I$ on the properties of a finite beam}

Eq. (\ref{field vector of beam in MM}) together with Eq. (\ref{MM in cylindrical system}) in the
circular cylindrical system is an exact beam solution to the Maxwell equations in free space. It
does not rely on the paraxial condition, and the field scalar $f=f({k_{\rho},\varphi})$ can be any
physically allowed function of $k_{\rho}$ and $\varphi$. In this section we are concerned with the
impact of angle $\theta_I$ on the properties of a beam and therefore consider only such a field
scalar $f=f(k_{\rho})$ that is $\varphi$-independent, sharply peaked at $k_{\rho}=0$, and
square-integrable. Furthermore, due to the relation $\mathbf{I} (\theta_I+ \pi)=
-\mathbf{I}(\theta_I)$, it is enough to assume $|\theta_I| \le \frac{\pi}{2}$. In the following,
the $\theta_I$ will be confined to this interval.

\subsection{Field vector distribution in the first-order approximation}

A sharply peaked field scalar $f(k_\rho)$ around $k_{\rho}=0$ means that the half divergence angle
$\Delta \theta$ of the beam satisfies
\begin{equation} \label{paraxial condition}
\Delta \theta \ll \frac{1}{2 \pi}.
\end{equation}
Upon considering integral (\ref{MM in cylindrical system}), $\frac{k_{\rho}}{k}$ in the MM can be
regarded as a small number in comparison with unity. So the elements of the MM can be expanded as
a power series in $\frac{k_{\rho}}{k}$. When $\theta_I$ is either equal to $\frac{\pi}{2}$
\cite{Lax-LM} or to $0$ \cite{Li1}, the lowest correction to the zeroth-order transverse component
is a second-order term \cite{remark}. In this subsection, I will show that when $\theta_I$
satisfies
\begin{equation} \label{requirement of thetaI}
\Delta \theta \ll |\theta_I| \leq \frac{\pi}{2},
\end{equation}
the transverse component will have a $\theta_I$-dependent first-order correction.

We rewrite Eq. (\ref{normalization factor}) as
$$
|\mathbf{k} \times \mathbf{I}|= |k_z \sin \theta_I| \left( 1- 2 \frac{k_x}{k_z} \cot \theta_I+
 \frac{k_x^2 \cos^2 \theta_I+ k_y^2}{k_z^2 \sin^2 \theta_I} \right)^{1/2}.
$$
The condition (\ref{requirement of thetaI}) guarantees that the second and the third parts in the
second factor are the first-order and the second-order terms in comparison with the first part. As
a result, in the first-order approximation, we have for the MM,
\begin{equation} \label{MM at 1st order approx}
m \approx m_0+ m_1,
\end{equation}
where
$$
m_0=  \mathrm{sgn}(\theta_I) \left(
                                   \begin{array}{cc} 1 & 0 \\
                                                 0 & 1 \\
                                                 0 & 0
                                             \end{array}
                                   \right)
$$
is the zeroth-order term,
$$
m_1= \mathrm{sgn}(\theta_I)
  \left(
        \begin{array}{cc}
                 0                         &  \cot \theta_I \sin\varphi \\
               -\cot \theta_I \sin\varphi  &   0 \\
               -\cos\varphi                & -\sin\varphi
        \end{array}
  \right) \frac{k_{\rho}}{k}
$$
is the first-order correction, and $\mathrm{sgn}$ is the sign function. Substituting Eq. (\ref{MM
at 1st order approx}) into Eqs. (\ref{field vector of beam in MM}) and (\ref{MM in cylindrical
system}) yields the field vector,
\begin{equation} \label{field vector of 1st approx}
\mathbf{F}(\mathbf{x}) \approx \mathbf{F}_T^{(0)}(\mathbf{x})+ \mathbf{F}_T^{(1)}(\mathbf{x})+
\mathbf{F}_L^{(1)}(\mathbf{x}),
\end{equation}
where
\begin{equation} \label{field vector of 0th approx}
\mathbf{F}_T^{(0)}(\mathbf{x})= m_0 \tilde{\alpha} F_0(r,z)
\end{equation}
is the zeroth-order term that is transverse and uniformly polarized,
\begin{equation} \label{1st correction to tc}
\mathbf{F}_T^{(1)}(\mathbf{x})= i \mathrm{sgn}(\theta_I) (\alpha_s \mathbf{e}_x- \alpha_p
\mathbf{e}_y) \frac{y}{r} F_1(r,z) \cot \theta_I
\end{equation}
is the first-order correction to the transverse component which is also uniformly polarized but is
$\theta_I$-dependent,
\begin{equation} \label{longitudinal field}
\mathbf{F}_{L}^{(1)}(\mathbf{x})= -i \mathrm{sgn}(\theta_I) \mathbf{e}_z \frac{\alpha_p x+
\alpha_s y}{r} F_1(r,z) \equiv F_{L}^{(1)} \mathbf{e}_z
\end{equation}
is the longitudinal component,
$$
F_0(r,z)= \int_0^k f(k_{\rho}) \exp(i k_z z) J_0 (r k_{\rho}) k_{\rho} dk_{\rho},
$$
and
$$
F_1(r,z)= \int_0^k \frac{k_{\rho}}{k} f(k_{\rho}) \exp(i k_z z) J_1 (r k_{\rho}) k_{\rho}
dk_{\rho}.
$$
In deriving Eq. (\ref{field vector of 1st approx}), I have made use of the following expansion,
\begin{equation} \label{expansion}
\exp(i \rho \cos \psi)= \sum_{m=-\infty}^{\infty} i^m J_m (\rho) \exp(im \psi),
\end{equation}
where $J_m$'s are the Bessel functions of the first kind.

It is noticed that the polarization state of the first-order transverse term is different from
that of the zeroth-order transverse term. As a matter of fact, they are orthogonal to each other.
Combining Eqs. (\ref{field vector of 0th approx}) and (\ref{1st correction to tc}) together, we
get the transverse component of the electric-field vector,
\begin{equation} \label{transverse component}
\mathbf{F}_{T}(\mathbf{x})= \mathrm{sgn}(\theta_I) \left[(\alpha_p F_0+ i \alpha_s \frac{y}{r} F_1
\cot \theta_I) \mathbf{e}_x+ (\alpha_s F_0- i \alpha_p \frac{y}{r} F_1 \cot \theta_I) \mathbf{e}_y
\right].
\end{equation}
This shows that the transverse component of the beam is not in general uniformly polarized. The
local polarization state is dependent on the value of $\theta_I$.

When $|\theta_I|= \frac{\pi}{2}$, the first-order correction to the transverse component vanishes,
$\mathbf{F}_T^{(1)} (\mathbf{x})=0$. In this case the zeroth-order term of the transverse
component is the field vector of uniformly-polarized beams, the polarization state being the same
as that of the angular spectrum, $\tilde{\alpha}$. It satisfies, together with the first-order
longitudinal component (\ref{longitudinal field}), the approximate transversality condition
\cite{Lax-LM},
\begin{equation} \label{approx TC}
\nabla_T \cdot \mathbf{F}_T^{(0)}+ i k F_{L}^{(1)}= 0,
\end{equation}
where $\nabla_T= \mathbf{e}_x \frac{\partial}{\partial x}+ \mathbf{e}_y \frac{\partial}{\partial
y}$.

Defining $m_c=\frac{\alpha_s}{\alpha_p}$ and choosing $\alpha_p= -\frac{1}{(1+|m_c|^2)^{1/2}}$ by
use of the normalization condition (\ref{normalization condition}), we find for the unit field
vector of the angular spectrum,
\begin{equation} \label{unit vector of angular spectrum}
\frac{\mathbf{f}}{|\mathbf{f}|}= -\mathrm{sgn}(\theta_I) \frac{(1+m_c\frac{k_y}{k} \cot\theta_I)
\mathbf{e}_x+ (m_c-\frac{k_y}{k} \cot\theta_I) \mathbf{e}_y- (\frac{k_x}{k}+ m_c \frac{k_y}{k})
\mathbf{e}_z}{(1+|m_c|^2)^{1/2}}.
\end{equation}
When $\theta_I= -\theta$ with $\theta>0$, Eq. (\ref{unit vector of angular spectrum}) turns into
\begin{equation} \label{unit vector of angular spectrum-1}
\frac{\mathbf{f}}{|\mathbf{f}|}= \frac{(1-m_c\frac{k_y}{k} \cot\theta) \mathbf{e}_x+
(m_c+\frac{k_y}{k} \cot\theta) \mathbf{e}_y-(\frac{k_x}{k}+ m_c \frac{k_y}{k})
\mathbf{e}_z}{(1+|m_c|^2)^{1/2}},
\end{equation}
which is the same as the equation (20) of Ref. \cite{Bliokh-B2} if $\theta$ is interpreted as the
incidence angle. This shows that the incident beam of Ref. \cite{Onoda-MN1} can be described in
this representation formalism as the first-order approximation of such a special beam the unit
vector $\mathbf{I}$ of which happens to make an angle of the minus incidence angle with the
propagation axis and happens to be normal to the interface. This is implied in Ref.
\cite{Onoda-MN1} by the unit vector $\mathbf{n}$ that plays the role of $\mathbf{I}$ and explains
why the physical properties of the incident beam in Ref. \cite{Onoda-MN1} depend on the
``incidence angle'' \cite{Bliokh-B2}. A similar incident beam was once proposed by Schilling
\cite{Schilling}.

Furthermore, when $\theta= \frac{\pi}{2}$, Eq. (\ref{unit vector of angular spectrum-1}) reduces
to
\begin{equation} \label{unit vector of angular spectrum-2}
\frac{\mathbf{f}}{|\mathbf{f}|}=\frac{\mathbf{e}_x+ m_c \mathbf{e}_y- (\frac{k_x}{k}+ m_c
\frac{k_y}{k}) \mathbf{e}_z} {(1+|m_c|^2)^{1/2}}.
\end{equation}
Upon noticing that $m_c= \frac{f_y}{f_x}$ in this case, Eq. (\ref{unit vector of angular
spectrum-2}) is exactly the same as the equation (22) of Ref. \cite{Bliokh-B2}. This shows that
the incident beam of Ref. \cite{Bliokh-B1} is nothing but the fundamental Gaussian beam, as long
as the first-order longitudinal component is taken into account.

\subsection{Transverse effect}

Now we are ready to discuss a transverse effect. When $|\theta_I|$ is not equal to
$\frac{\pi}{2}$, the first-order term of the transverse component does not vanish. Since this term
is not axisymmetric as is clearly shown by Eq. (\ref{1st correction to tc}), its interference with
the zeroth-order term renders the intensity distribution deformed from the axisymmetry as can be
seen from Eq. (\ref{transverse component}). A direct consequence of this deformation is the
following transverse effect: the barycenter of a beam of elliptically polarized angular spectrum
is displaced from the center in the transverse $y$-direction; the displacement is dependent on the
value of $\theta_I$ and on the polarization ellipticity of the angular spectrum.

To show this, let us define the $y$-position $y_b$ of the beam's barycenter as the expectation of
the $y$-coordinate of the beam,
\begin{equation} \label{definition of barycenter}
y_b= \langle y \rangle= \frac{\int \int \mathbf{F}^{\dag} y \mathbf{F} dx dy}{\int \int
\mathbf{F}^{\dag} \mathbf{F} dx dy},
\end{equation}
where superscript $\dag$ stands for the conjugate transpose. According to Eq. (\ref{electric field
of a beam}), we have
$$
\int \int \mathbf{F}^{\dag} y \mathbf{F} dx dy= \int \int (i \mathbf{f}^{\dag} \frac{\partial
\mathbf{f}} {\partial k_y}+ z \frac{k_y}{k_z} \mathbf{f}^{\dag} \mathbf{f} ) dk_x dk_y
$$
and
$$
\int \int \mathbf{F}^{\dag} \mathbf{F} dx dy= \int \int \mathbf{f}^{\dag} \mathbf{f} dk_x dk_y.
$$
Furthermore, Eqs. (\ref{factorization}) and (\ref{separation}) tell us that
$$
\mathbf{f}^{\dag} \mathbf{f}= |f|^2
$$
and
$$
\mathbf{f}^{\dag} \frac{\partial \mathbf{f}} {\partial k_y}= f^* \frac{\partial f} {\partial k_y}+
\tilde{\alpha}^{\dag} m^T \frac{\partial m} {\partial k_y} \tilde{\alpha} |f|^2.
$$
Substituting all these into Eq. (\ref{definition of barycenter}) and noticing that $f$ is an even
function of $k_y$, we obtain
\begin{equation}
y_b= i \frac{\int \int \tilde{\alpha}^{\dag} m^T \frac{\partial m} {\partial k_y} \tilde{\alpha}
|f|^2 dk_x dk_y}{\int \int |f|^2 dk_x dk_y},
\end{equation}
which shows that the $y$-position of the beam's barycenter is independent of $z$. Substituting Eq.
(\ref{form of MM}) and $\tilde{\alpha}$, we get
\begin{equation} \label{transverse displacement}
y_b=-\frac{\sigma \cot \theta_I}{2 \int_0^k |f(k_{\rho})|^2 k_{\rho} dk_{\rho}}
    \int_0^k \left(1+ \frac{k_z-k \cos \theta_I}{|k_z-k \cos \theta_I|} \right) \frac{|f(k_{\rho})|^2}{k_z} k_{\rho}
    dk_{\rho},
\end{equation}
where $\sigma= -i(\alpha_p^* \alpha_s-\alpha_p \alpha_s^*)$ is the polarization ellipticity of the
angular spectrum. It is emphasized that this expression has a validity that does not depend on the
paraxial condition (\ref{paraxial condition}) and large angle condition (\ref{requirement of
thetaI}). It is inferred from the expression that

\noindent (1) the beam is indeed displaced a distance transversely from the center, because $y_b$
does not change on propagation;

\noindent (2) the opposite ellipticity $\sigma$ corresponds to the opposite displacement for a
given $\theta_I$;

\noindent (3) the opposite angle $\theta_I$ corresponds to the opposite displacement for a given
$\sigma$.

Let us now discuss the dependence of the transverse displacement on the angle $\theta_I$. If
$|\theta_I|=\frac{\pi}{2}$, the first factor $\cot \theta_I$ of Eq. (\ref{transverse
displacement}) tells us that the displacement is equal to zero. This is the case that corresponds
to the uniformly polarized beams (\ref{field vector of 0th approx}) in the zeroth-order
approximation. If $\theta_I=0$, the second term in the integrand cancels the first one, also
resulting in zero displacement. In fact, this case corresponds to the cylindrical vector beams
which are axially symmetric in both polarization and intensity distribution \cite{Li1}. For a
$\theta_I$ that is neither equal to $\frac{\pi}{2}$ nor equal to $0$, Eq. (\ref{transverse
displacement}) is rewritten as

\begin{equation} \label{transverse displacement 1}
y_b=-\frac{\sigma \cot \theta_I}{2 \int_0^k |f|^2 k_{\rho} dk_{\rho}}
    \left(
       \int_0^k \frac{|f|^2}{k_z} k_{\rho} dk_{\rho}
      +\int_0^{k \sin|\theta_I|} \frac{|f|^2}{k_z} k_{\rho} dk_{\rho}
      -\int_{k \sin|\theta_I|}^k \frac{|f|^2}{k_z} k_{\rho} dk_{\rho}
    \right).
\end{equation}
If $|\theta_I| \gg \Delta \theta$, the third integral is much smaller than the second one,
remembering that $f(k_{\rho})$ is a sharply peaked function about $k_{\rho}=0$. As a result, the
transverse displacement in this case becomes
$$
y_b \approx -\frac{\sigma \cot \theta_I}{\int_0^k |f|^2 k_{\rho} dk_{\rho}} \int_0^k
\frac{|f|^2}{k_z} k_{\rho} dk_{\rho}.
$$
Under the paraxial condition which means $k_z \approx k$ for the denominator of the integrand in
the first-order approximation with respect to $\frac{k_{\rho}}{k}$, it reduces to
\begin{equation}\label{TD at large angle}
y_b \approx -\frac{\sigma}{k} \cot \theta_I.
\end{equation}
Due to the factor $\cot\theta_I$ in Eq. (\ref{transverse displacement 1}), the smaller
$|\theta_I|$ goes, the larger the magnitude of the displacement becomes until the following
relation establishes,
\begin{equation}\label{cancellation}
\int_0^{k \sin|\theta_I|} \frac{|f|^2}{k_z} k_{\rho} dk_{\rho}= \int_{k \sin|\theta_I|}^k
\frac{|f|^2}{k_z} k_{\rho} dk_{\rho}.
\end{equation}
At this point, the third integral of Eq. (\ref{transverse displacement 1}) cancels the second one,
and the displacement takes the form of
\begin{equation}\label{maximum displacement}
y_b=-\frac{\sigma \cot \theta_{I0}}{2 \int_0^k |f|^2 k_{\rho} dk_{\rho}}
    \int_0^k \frac{|f|^2}{k_z} k_{\rho} dk_{\rho},
\end{equation}
where $\theta_{I0}$ is the solution to Eq. (\ref{cancellation}). Similarly it reduces to, under
the paraxial condition,
$$
y_b \approx -\frac{\sigma \cot \theta_{I0}}{2 k}.
$$
If $|\theta_I|$ goes even smaller, the third integral outstrips the second one in magnitude, and
the displacement becomes smaller in magnitude until $\theta_I=0$ when the displacement is equal to
zero. It is thus expected that Eq. (\ref{maximum displacement}) represents the maximum
displacement if only the change of $\theta_I$ is considered. Roughly speaking, the value of
$|\theta_{I0}|$ that satisfies Eq. (\ref{cancellation}) is approximately equal to the half
divergence angle of the beam, $|\theta_{I0}| \sim \Delta \theta= \frac{1}{k w_0}$, where $w_0$ is
the half width of the beam. For a paraxial beam in which $\Delta \theta$ is very small, the
maximum transverse displacement can be as large as the order of $w_0$ \cite{Li3}, $|y_b| \approx
\frac{w_0}{2}$, for circularly polarized angular spectra $\sigma= \pm 1$.

It is interesting to note that Schilling \cite{Schilling} once found the transverse displacement
more than 40 years ago for an incident beam the unit field vector of which can be described by Eq.
(\ref{unit vector of angular spectrum-1}). Recently, Bliokh and Bliokh \cite{Bliokh-B2}
rediscovered Schilling's result when comparing the incident beams of Refs. \cite{Onoda-MN1} and
\cite{Bliokh-B1}.

\section{\label{conclusions} Conclusions and Remarks}

In conclusion, I factorized in Eq. (\ref{factorization}) the field vector of a beam's angular
spectrum into the MM and the Jones-like vector and showed that the degree of freedom of the MM can
be described by the azimuthal angle of a fixed unit vector with respect to the wave vector. This
degree of freedom provides us with such a beam solution in which every plane-wave component is
specified by the same fixed unit vector $\mathbf{I}$ and has the same normalized Jones-like vector
$\tilde{\alpha}$. The integral representation for the MM of a beam's field vector was formulated
in Eq. (\ref{MM of beam}) [or (\ref{MM in cylindrical system}) in the circular cylindrical system]
by letting the unit vector lie in the plane $zox$ and make an angle $\theta_I$ with the $z$-axis.

The impact of the angle $\theta_I$ was discussed on the vectorial property of a beam that has a
$\varphi$-independent field scalar $f(k_{\rho})$. The electric-field vector (\ref{field vector of
1st approx}) was obtained in the first-order approximation under the paraxial condition
(\ref{paraxial condition}) for large $|\theta_I|$ that satisfies Eq. (\ref{requirement of
thetaI}). It was shown that the transverse component has a $\theta_I$-dependent first-order
correction. This is different from the cases of $|\theta_I|= \frac{\pi}{2}$ and $\theta_I=0$ in
which the lowest correction to the zeroth-order transverse component is a second-order term. A
transverse effect was found and the dependence of the transverse displacement on $\theta_I$ was
discussed. The paraxial approximation of the transverse displacement was also given. In a word,
the angle $\theta_I$ in the representation formalism advanced here plays the role of a parameter
that describes the vectorial property of the beam.

A physically allowed field scalar $f(k_{\rho},\varphi)$ in Eq. (\ref{MM in cylindrical system})
can be expanded as a Fourier series,
\begin{equation}
f(k_{\rho},\varphi)= \sum_{l=-\infty}^{\infty} f_l (k_{\rho}) \exp(i l \varphi).
\end{equation}
One may consider the constituent term of the following form,
\begin{equation} \label{constituent amplitude scalar}
f(k_{\rho},\varphi)= f_l (k_{\rho}) \exp(i l \varphi),
\end{equation}
and discuss the impact of angle $\theta_I$ on the resultant beam. It is expected that when
$\theta_I= \pm \frac{\pi}{2}$, Eq. (\ref{constituent amplitude scalar}) together with Eqs.
(\ref{field vector of beam in MM}) and (\ref{MM in cylindrical system}) will yield the eigen beam
of the orbital angular momentum \cite{Allen-BSW} in the zeroth-order approximation.

According to the triad relation expressed by the first equation of (\ref{triad}) and the principle
of duality in free space \cite{Davis-P1}, I would like to point out that the following $3 \times
2$ matrix
\begin{equation}
m_M=      \left(
                \begin{array}{cc}
                       s_x & -p_x \\
                       s_y & -p_y \\
                       s_z & -p_z
                \end{array}
          \right)
\end{equation}
can be regarded as the MM for the magnetic-field vector, which maps a Jones-like vector to the
3-component magnetic-field vector for a particular wave vector.

The representation formalism of finite electromagnetic beams developed in this paper depends
closely on the MM and its degree of freedom. It is worth noting that in this representation
formalism, the Jones-like vector does not depend on the MM's degree of freedom. Therefore one
Jones-like vector can be mapped to an infinite number of field vectors due to the MM's degree of
freedom, as can be seen from Eq. (\ref{factorization}) as well as Eq. (\ref{field vector of beam
in MM}). The physical significance of the MM's degree of freedom needs further investigation.

\section*{Acknowledgments}
The author would like to thank Franco Gori, Thomas G. Brown, and Masud Mansuripur for their
helpful discussions. This work was supported in part by the National Natural Science Foundation of
China (60877055 and 60806041), the Science and Technology Commission of Shanghai Municipal
(08JC14097 and 08QA14030), the Shanghai Educational Development Foundation (2007CG52), and the
Shanghai Leading Academic Discipline Program (T0104).


\begin{thebibliography}{99}
\bibitem{remark} If $\tilde{\alpha}= \left( \begin{array}{c} 0 \\ 1 \end{array} \right)$ which corresponds to the TE beam mode
\cite{Davis-P1}, the transverse component has only the zeroth-order term \cite{Li1}.
\end{thebibliography}
\end{document}